\documentclass[aps,prl,10pt,twoside,reprint,superscriptaddress,nofootinbib,floatfix]{revtex4-2}
\usepackage[utf8]{inputenc} 
\usepackage[T1]{fontenc}    
\usepackage{hyperref}       
\usepackage{url}            
\usepackage{amsfonts}       
\usepackage{nicefrac}       
\usepackage{microtype}      
\usepackage{graphicx}
\usepackage{amsmath}
\usepackage{float}
\usepackage[capitalize]{cleveref} 
\usepackage[table]{xcolor}
\usepackage{color}
\usepackage{cleveref}
\usepackage{colortbl}
\usepackage{multirow}
\usepackage{tikz}
\usepackage{soul}
\usepackage{bm}
\usepackage{xfrac}

\begin{document}
\title{RHIC \texorpdfstring{$\sqrt{s_{NN}}=200$}{√sₙₙ = 200} GeV hadron yields  and the isospin dependent equation of state}

\author{Feyisola Nana}
\affiliation{Illinois Center for Advanced Studies of the Universe, Department of Physics, University of Illinois at Urbana-Champaign, Urbana, IL 61801, USA}
\author{Jordi Salinas~San~Mart\'in}
\affiliation{Illinois Center for Advanced Studies of the Universe, Department of Physics, University of Illinois at Urbana-Champaign, Urbana, IL 61801, USA}
\author{Jacquelyn Noronha-Hostler}
\affiliation{Illinois Center for Advanced Studies of the Universe, Department of Physics, University of Illinois at Urbana-Champaign, Urbana, IL 61801, USA}

\begin{abstract}
    The statistical hadronization model has been successful in extracting information at chemical freeze-out in heavy-ion collisions. 
    At RHIC, with a collision energy of $\sqrt{s_{NN}}=200$ GeV, many different ion species have been used for $A$+$A$ collisions.
    This allows for a scan across the charge fraction $Y_Q=Z/A$, where $Z$ is the proton number and $A$ is the baryon number.
    We first make predictions for $A$+$A$ collisions that do not yet have published experimental data on hadron yield ratios (O+O, Ru+Ru, Zr+Zr). 
    We then use both the experimental and predicted yield ratios to perform thermal fits across $Y_Q$, enabling us to extract $s/n_B$ and other thermodynamic information at chemical freeze-out. 
    Using the relation between $s/n_B$ and $Y_Q$, we can calculate a new constraint on the finite temperature equation of state at finite densities. 
    We discuss implications of this constraint and propose future runs that can help connect to the equation of state relevant for neutron star mergers.
\end{abstract}

\maketitle
\newpage

{\it Introduction.}––Heavy-ion collisions at the relativistic heavy-ion collider (RHIC) explore the Quantum Chromodynamics (QCD) phase diagram at high temperatures $T$ and a range of baryon densities $n_B$.
In the early stages of these collisions, the quark gluon plasma (QGP) is produced. It then expands and cools until the quarks and gluons within the QGP eventually hadronize, and the newly produced particles freeze-out. 
At the point of freeze-out, it is assumed that the fireball is at chemical equilibrium such that one can use a statistical hadronization model (SHM) \cite{Dashen:1969ep,Cleymans:1999st,Braun-Munzinger:2001hwo,PBM,Becattini:2001fg,Andronic:2005yp,Cleymans:2005xv,Wheaton:2011rw,Andronic:2017pug,Andronic:2017pug,Andronic:2018qqt,Andronic:2021dkw,Cleymans:2020aqh,Vovchenko:2019pjl} to extract the temperature and chemical potentials for a specific beam energy $\sqrt{s_{NN}}$ and colliding species containing $A$ nucleons and $Z$ protons using so-called thermal fits.

From QCD there are three conserved charges in heavy-ion collisions: baryon number (B), strangeness (S), and electric charge (Q).  
While local fluctuations of BSQ charge densities exist \cite{Carzon:2019qja,Plumberg:2024leb}, globally the system remains strange-neutral throughout the evolution such that the global \textit{average} strangeness density is zero,
\begin{equation}\label{Eq:strangeness_neutrality}
    \langle n_S\rangle =0.
\end{equation}
Since both electric charge and baryon number are conserved, the ratio of charge fraction (average electric charge density over the average baryon density)
\begin{equation}\label{Eq:YQ_conservation}
    Y_Q \equiv \frac{Z}{A} = \frac{\langle n_Q\rangle }{\langle n_B\rangle}
\end{equation}
is globally conserved as well. 
Thus, given a specific ion species, $Y_Q$ is defined for the system and conserved throughout the entire evolution. 
To date, experiments at RHIC have covered a wide set of species with a significant range of $Y_Q=[0.387,0.5]$ for a collision energy of $\sqrt{s_{NN}}=200$ GeV, as shown in Table~\ref{Table:YQ_by_system} \cite{STAR:2010dor,STAR:2008med,STAR:2022nvh}.
Equations\ (\ref{Eq:strangeness_neutrality}-\ref{Eq:YQ_conservation}) define the {\it strangeness neutrality} and \textit{charge fraction conservation} conditions. 
By applying these conditions, the phase space of $\left\{T,\mu_B,\mu_S,\mu_Q\right\}$ can be constrained so that our equation of state simplifies to a function of just $\left\{T,\mu_B\right\}$. 
The other chemical potentials are then functions of $T$ and $\mu_B$, specifically $\mu_S = \mu_S(T,\mu_B)$ and  $\mu_Q=\mu_Q(T,\mu_B)$. 
With these constraints, one can calculate thermodynamic quantities at the point of freeze-out, such as energy density $\varepsilon$ and entropy over the baryon density, $s/n_B$ \cite{Ejiri:2005uv,Alba:2014eba,Guenther:2017hnx,Ratti:2018ksb}.

\renewcommand{\thempfootnote}{\fnsymbol{mpfootnote}}
\begin{table}[h!] 
\centering
\begin{tabular}{l@{\hspace{18pt}}c@{\hspace{18pt}}c@{\hspace{18pt}}c@{\hspace{18pt}}c} 
    \toprule
    System      & $Z$   & $A$   & $Y_Q$ & Published yield data? \\
    \colrule
    O+O         & 8     & 16    & 0.500 & no \\
    Cu+Cu       & 29    & 63    & 0.460 & yes \\ 
    Ru+Ru       & 44    & 96    & 0.458 & no\footnote{Preliminary data has recently been presented in Ref.~\cite{Ma:2024wwnd}.} \\
    Zr+Zr       & 40    & 96    & 0.417 & no\footnotemark[1] \\
    Au+Au       & 79    & 198   & 0.399 & yes  \\
    U+U         & 92    & 238   & 0.387 & yes  \\
    \botrule
\end{tabular}
\caption{Collision systems for heavy-ion experiments at approximately $\sqrt{s_{NN}}=200$ GeV with their corresponding proton number, atomic mass, charge fraction, and data availability status for hadron yield ratios of identified charged particles.
}\label{Table:YQ_by_system}
\end{table}

While heavy-ion collisions probe (approximately) symmetric nuclear matter, neutron star mergers can only be described via isospin dependent equations of state \cite{Brandt:2017oyy,Aryal:2020ocm,Choi:2020eun,Wen:2020nqs,Carlomagno:2021gcy,Drischler:2021kxf,Lopes:2021tro,Liu:2021uoz,Alford:2022bpp,Alford:2023rgp,Brandt:2022hwy,Most:2022wgo,Ayala:2023cnt,Ayala:2024sqm,Andersen:2023ofv,Andersen:2023ivj,Sorensen:2023zkk,MUSES:2023hyz,Routray:2024kgv,Kojo:2024sca}. It has recently been pointed out in Ref. \cite{Mroczek:2024sfp} that the dense matter equation of state, can be explored through a well-defined expansion up to second order in $T$ for $(T/\mu_B)\ll 1$.
This expansion can be cast in terms of derivatives of $s/n_B$ with respect to temperature and the isospin asymmetry parameter defined by
\begin{equation}\label{Eq:delta_definition}
    \delta \equiv 1-2Y_Q,
\end{equation}
namely,
\begin{equation*}
    \left.\frac{\partial (s/n_B)}{\partial T}\right\lvert_{\delta=0,\,T=0},\quad
    \left.\frac{\partial^3 (s/n_B)}{\partial \delta^2 \partial T}\right\lvert_{\delta=0,\,T=0}.
\end{equation*}
The number of different colliding systems that have been run at RHIC at $\sqrt{s_{NN}}=200$ GeV open the possibility to extract the derivative 
\begin{equation}\label{Eq:d2snB_ddelta2}
    \left.\frac{\partial^2 (s/n_B)}{\partial \delta^2 }\right\lvert_{\delta=0,\,T=\text{const.}},
\end{equation}
from heavy-ion data at a fixed point in $\left\{T,\mu_B,\mu_S\right\}$ ($\mu_Q$, of course, varying with $\delta$). 
In principle, if multiple ions were run at other beam energies, we could then extract the dependence of this derivative on $T,\mu_B$, and $\mu_S$ as well. 
Currently, hadron yield data does not yet exist for all the species listed in Table\ \ref{Table:YQ_by_system}.  
Extensive data exist for Au+Au \cite{STAR:2008med} at $\sqrt{s_{NN}}=200$ GeV for $0$--$5\%$ centrality, with a smaller subset of data available for U+U \cite{STAR:2022nvh} at $\sqrt{s_{NN}}=197$ GeV for $0$--$5\%$ centrality, and for Cu+Cu  \cite{STAR:2010dor} at $\sqrt{s_{NN}}=200$ GeV for $0$--$10\%$ centrality.
To be consistent across systems, we limit ourselves to yield ratios of identified charged particles where data exist for all three ion pairs. 
While Zr+Zr, Ru+Ru, and O+O have all been run experimentally, the hadron yield ratio data have not yet been published for these species. 

In this Letter, we first develop a method to predict the hadron yield ratios of these species based on the available knowledge from Cu+Cu, Au+Au, and U+U.
Then, using a combination of our predicted hadron yield ratios and the existing ratios, we are able to calculate the derivative in Eq.~(\ref{Eq:d2snB_ddelta2}) and comment on its implication for the dense matter equation of state. 
We confirm that $T$, $\mu_B$, and $\mu_S$ are nearly constant for a fixed collision energy while we vary $Y_Q$. 
We then determine that the derivative is $\frac{\partial^2 (s/n_B)}{\partial \delta^2 }\big|_{\delta=0,\,T\approx 145\, \text{MeV}}=-0.183_{-0.166}^{+0.128}\times10^5$.
We propose future runs that can provide further insight into the dense matter equation of state relevant to neutron star mergers.

{\it Estimating hadron yield ratios.}––One of the most successful statistical models used to describe strongly interacting matter in the hadronic phase is the Hadron Resonance Gas (HRG) model. 
This model represents hadrons using the grand canonical ensemble (GCE) framework. 
The basic quantity required to compute thermodynamic properties and the thermal composition of the particle (hadron) yields measured in relativistic heavy-ion collisions is the partition function $\mathcal{Z}(T, V)$ \cite{PBM}. 
Within a strongly interacting medium where the three conserved charges are included, the GCE partition function can be written as a sum over all known hadrons and resonances listed in the Review of Particle Physics by the Particle Data Group (PGD),
\begin{equation}
\ln \mathcal{Z}_{\text{HRG}}(T, V, \vec{\mu})=\sum_{i \in \text{PDG}} \ln \mathcal{Z}_i(T, V, \vec{\mu}),
\end{equation}
where
\begin{equation}\label{Eq:particle_partition_function}
\ln \mathcal{Z}_i(T, V, \vec{\mu}) =\pm\frac{g_i V}{2 \pi^2} \int_0^{\infty} p^2 \mathrm{~d} p \ln \left[1 \pm e^{-(E_i-\mu_i)/T}\right]
\end{equation}
is the partition function of the $i$-th particle, $\vec{\mu} = (\mu_B, \mu_S, \mu_Q)$ represents the chemical potentials associated with the conserved charges, $\mu_i=(B_i \mu_B + S_i \mu_S + Q_i \mu_Q)$, the total energy is $E_i=\sqrt{{p^2}+m_i^2}$ for a particle with mass $m_i$, and $g_i = (2J_i+1)$ is the spin degeneracy factor. 
The positive sign corresponds to fermions, and the negative sign to bosons. 
By performing the momentum integration in Eq.~\ref{Eq:particle_partition_function}, one can analytically determine the total pressure as well as other thermodynamic quantities such as the particle number density $n_i$, entropy density $s_i$, and energy density $\varepsilon_i$:
\begin{widetext}
\begin{equation}
n_i \left(T, \mu_{i}\right) = \frac{T}{V} \left( \frac{\partial \ln \mathcal{Z}_i}{\partial \mu_i} \right)_{T, V} = \frac{g_i}{2 \pi^2} \int_0^{\infty} \frac{p^2 \, \mathrm{d} p}{\exp \left[ (E_i - \mu_i)/{T} \right] \pm 1},
\end{equation}
\begin{equation}
\varepsilon_i \left(T, \mu_{i}\right) =-\frac{1}{V}\left(\frac{\partial \ln \mathcal{Z}_i}{\partial(1 / T)}\right)_{\mu_i / T, V} = \frac{g_i}{2 \pi^2} \int_0^{\infty} \frac{p^2 E_i \, \mathrm{d} p}{\exp \left[ (E_i - \mu_i)/T \right] \pm 1} ,
\end{equation}
\begin{equation}
s_i \left(T, \mu_{i}\right) = \frac{1}{V} \left( \frac{\partial \left(T \ln \mathcal{Z}_i \right)}{\partial T} \right)_{\mu_i, V}
= \pm \frac{g_i}{2 \pi^2} \int_0^{\infty} p^2 \mathrm{~d} p\left(\ln \left(1 \pm \exp \left[-\left(E_i-\mu_i\right) / T\right]\right) \pm \frac{E_i-\mu_i}{T\left(\exp \left[\left(E_i-\mu_i\right) / T\right] \pm 1\right)}\right).
\end{equation}
\end{widetext}

\begin{table}[ht] 
\centering
\resizebox{\columnwidth}{!}{%
\begin{tabular}{l@{\hspace{4pt}}c@{\hspace{6pt}}c@{\hspace{6pt}}c}
    \toprule
    Ratio                       & Cu+Cu \cite{STAR:2010dor} & Au+Au \cite{STAR:2008med} & U+U \cite{STAR:2022nvh}   \\
    \colrule
    $\pi^-/\pi^+$               & $1.003 \pm 0.050$         & $1.015 \pm 0.051$         & $1.005 \pm 0.056$         \\
    $K^-/K^+$                   & $0.936 \pm 0.051$         & $0.965 \pm 0.048$         & $0.938 \pm 0.066$         \\
    $\bar{p}/p$ (uncorr.)       & $0.801 \pm 0.040$         & $0.769 \pm 0.055$         & $0.758 \pm 0.074$         \\
    $\bar{p}/p$ (corr.)         & $0.771 \pm 0.038$         & $0.740 \pm 0.053$         & $0.729 \pm 0.071$         \\
    $K^-/\pi^-$                 & $0.144 \pm 0.016$         & $0.151 \pm 0.018$         & $0.145 \pm 0.020$         \\
    $\bar{p}/\pi^-$ (uncorr.)   & $0.077 \pm 0.008$         & $0.082 \pm 0.010$         & $0.077 \pm 0.012$         \\
    $\bar{p}/\pi^-$ (corr.)     & $0.0489 \pm 0.0051$       & $0.0521 \pm 0.0063$       & $0.0489 \pm 0.0076$       \\
    \botrule
\end{tabular}
}
\caption{Experimentally measured hadron yield ratios and corresponding error of identified charged particles in Cu+Cu, Au+Au, and U+U collision systems. STAR collaboration data at $\sqrt{s_{NN}}=200$ GeV for Cu+Cu and Au+Au; in the case of U+U, data from the $\sqrt{s_{NN}}=197$ GeV run is provided.
Both corrected and uncorrected results for feed-down contributions are shown.
}
\label{Table:experimental_yield_ratios}
\end{table}

We study the following ratios of hadronic yields: $\pi^{-}/\pi^{+}$, $K^{-}/K^{+}$, $\Bar{p}/p$, $K^{-}/\pi^{-}$, and $\Bar{p}/\pi^{-}$ rather than yields, to minimize system size effects\footnote{Data also exist for $K^+/\pi^+$ and $p/\pi^+$, but these are redundant given the other ratios available, so we exclude it from our fits.}. 
The summary of the ratios used and their experimental values can be found in Table~\ref{Table:experimental_yield_ratios}. 
Additionally, some of the systematic uncertainties cancel when studying ratios (as compared to yields), since both particles in each ratio are detected under the same experimental conditions. 
Most of the available experimental data do not account for the contribution of feed-down from weak decays, mainly affecting the (anti)proton yields. 
Following the estimations from Refs.~\cite{Andronic:2005yp,Flor:2020fdw}, our model corrects for this contribution by scaling the proton and antiproton yields by $66\%$ and $63.5\%$, respectively, assuming an even correction across collision systems. 
Using these ratios, we perform a minimum $\chi^2$ fit using {\sc Thermal-FIST} \cite{Vovchenko:2019pjl} to obtain $\left\{T,\mu_B,\mu_S,\mu_Q\right\}$ at chemical freeze-out, given the strangeness neutrality condition and the electric-to-baryon charge ratio $Y_Q$.

To predict the yield ratios for species whose experimental runs have been completed but not yet published, we fit various functional forms to the available data, incorporating the error bars from existing experimental results. This approach allows us to estimate the yield ratios and associated uncertainties for O+O, Ru+Ru, and Zr+Zr systems as a function of $Y_Q$ by interpolating or extrapolating, as necessary. The fitted functional forms are polynomials of first-, second-, and third-orders:
\begin{equation}\label{Eq:polynomial_functions}
    p_m(Y_Q) = \sum_{i=0}^m a_{m,i}\,Y_Q^i\,; \quad \quad m=\{1,2,3\},
\end{equation}
where we vary the coefficients $a_{m,i}$ within a specified range (see Table~\ref{Table:YQ_polynomial_coefficients}) and eliminate fits that do not fall within the given error bars\footnote{The code used to perform this analysis can be found: \url{https://github.com/jordissm/YQ_InterpolatorAndExtrapolator.git}}. 

\begin{figure}[tbp!]
    \centering
    \includegraphics[width=0.9\columnwidth]{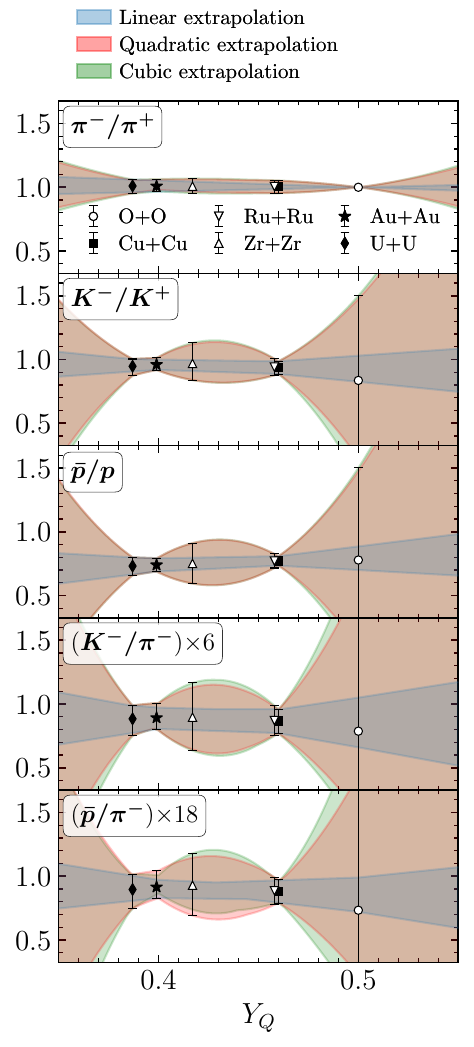}
    \caption{Charge fraction dependence of  hadron yield ratios. Experimental data points (closed symbols) for Cu+Cu \cite{STAR:2010dor}, Au+Au \cite{STAR:2008med}, U+U \cite{STAR:2022nvh} results and predicted ratios (open symbols)  for O+O, Ru+Ru, and Zr+Zr collision systems. The large error band for O+O is a result of its $Y_Q = 0.5$ being outside the regime of validity. $K^-/\pi^-$ are rescaled by $6$ and $\bar{p}/\pi^-$ by $18$. Feed-down contributions are included.
    }
    \label{Fig:predicted_yield_ratios}
 \end{figure}

In Fig.~\ref{Fig:predicted_yield_ratios} we show the results of our fits for each of the different yield ratios as functions of $Y_Q$.  Both the experimental data points (Cu+Cu, Au+Au, U+U) and predicted yield ratios (O+O, Ru+Ru, Zr+Zr) with errors are shown. 
In addition to the experimental constraints,
since O+O collisions should produce precisely symmetric nuclear matter ($Y_Q=0.5$), we set the $\pi^-/\pi^+$ yield ratio to exactly unity.
We summarize our results for the predicted yield ratios for O+O, Ru+Ru, and Zr+Zr in Table~\ref{Table:predicted_yield_ratios}. 

\begin{table}[tp!] 
\centering
\resizebox{\columnwidth}{!}{%
\begin{tabular}{l@{\hspace{4pt}}c@{\hspace{6pt}}c@{\hspace{6pt}}c}
    \toprule
    Ratio                       & O+O               & Ru+Ru             & Zr+Zr \\
    \colrule
    $\pi^-/\pi^+$               & $1.000$           & $1.003\pm0.051$   & $1.011\pm0.057$   \\
    $K^-/K^+$                   & $0.781\pm0.781$   & $0.941\pm0.065$   & $0.984\pm0.146$   \\
    $\bar{p}/p$ (uncorr.)       & $0.780\pm0.780$   & $0.801\pm0.057$   & $0.782\pm0.165$   \\
    $\bar{p}/p$ (corr.)         & $0.750\pm0.750$   & $0.771\pm0.055$   & $0.752\pm0.159$   \\
    $K^-/\pi^-$                 & $0.187\pm0.187$   & $0.145\pm0.020$   & $0.150\pm0.044$   \\
    $\bar{p}/\pi^-$ (uncorr.)   & $0.098\pm0.098$   & $0.077\pm0.009$   & $0.082\pm0.021$   \\
    $\bar{p}/\pi^-$ (corr.)     & $0.0625\pm0.0625$   & $0.0491\pm0.0060$   & $0.0520\pm0.0135$   \\
    \botrule
\end{tabular}
}
\caption{Predicted hadron yield ratios and corresponding error of identified charged particles in O+O, Ru+Ru, and Zr+Zr collision systems at $\sqrt{s_{NN}}=200$ GeV.
Both corrected and uncorrected results for feed-down contributions are shown.
}
\label{Table:predicted_yield_ratios}
\end{table}

\begin{figure*}[tbp!]
    \centering
    \includegraphics[width=0.9\textwidth]{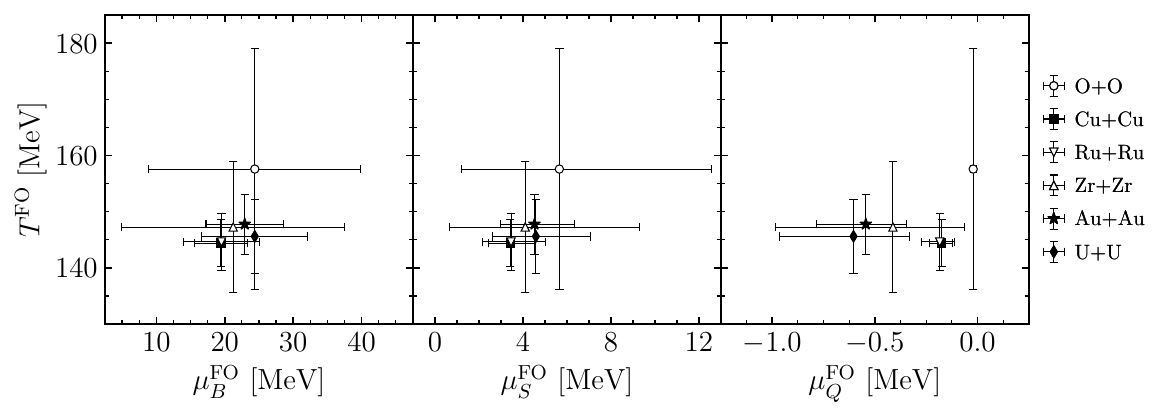}
     \caption{The QCD phase diagram  $T$–$\mu_{B,S,Q}$ planes at chemical freeze-out extracted from $\sqrt{s}=200$ GeV. The predicted ratios are marked by the open symbols while closed symbols indicate ratios of the fits from experimental data.
     }
     \label{Fig:phase_diagrams}
\end{figure*}

The predicted yield ratios for Ru+Ru and Zr+Zr are generally well estimated, with error bars comparable in size to the experimental data points (Ru+Ru is especially well constrained because its $Y_Q$ is nearly identical to Cu+Cu). 
Although the $K^-/\pi^-$ and $p/\pi^-$ ratios have slightly larger error bars, actual experimental measurements will help to reduce this uncertainty. 
However, O+O collisions have significantly larger uncertainties. 
The O+O error arises because its charge fraction, $Y_Q = 0.5$, falls outside the range of experimental $Y_Q$ values being considered $Y_Q =[0.387,0.460]$ and it is therefore difficult to extrapolate beyond the regime of validity.
In contrast, the $\pi^-/\pi^+$ yield ratio estimations have small uncertainties across the entire range of charge fractions considered, due to the additional constraint on symmetric nuclear matter.
Thus, of the three nuclei for which we have predicted yield ratios using the known experimental yield ratios, the experimental data for the yield ratios from O+O collisions should provide the most important constraints.

Now that we have yield ratios across the range $Y_Q=[0.387,0.5]$ from both the experimental data and theoretical predictions, we can use them to extract the thermodynamic information at chemical freeze-out. 
To do so, we use the hadron resonance gas model from the {\sc Thermal-FIST} \cite{Vovchenko:2019pjl} package. Within {\sc Thermal-FIST}, one can perform a minimum $\chi^2$ fit to the experimental yield ratios (see Appendix A for further details) using the most up-to-date hadronic list (PGD2021+) \cite{SanMartin:2023zhv}.
The output is then the freeze-out parameters $\left\{T^\text{FO},\mu_B^\text{FO},\mu_S^\text{FO},\mu_Q^\text{FO}\right\}$ for each $Y_Q$. 
Once $\left\{T^\text{FO},\mu_B^\text{FO},\mu_S^\text{FO},\mu_Q^\text{FO}\right\}$ along $Y_Q$ are known, we can then insert these values into the HRG calculations of the equation of state to determine the relevant thermodynamic quantities such as $s/n_B$, energy density $\varepsilon$, and the baryon density in terms of the saturation density $n_B/n_\text{sat}$.
We limit our analyses to the default thermal fits: ideal non-interacting gas model with quantum statistics, grand canonical ensemble, and energy-dependent Breit-Wigner resonance widths, while varying the thermal configurations for the different $Y_Q$ values. 

{\it Discussion.}––In Fig.~\ref{Fig:phase_diagrams} we show our resulting values for $\left\{T^\text{FO},\mu_B^\text{FO},\mu_S^\text{FO},\mu_Q^\text{FO}\right\}$ for different ion species, i.e., different values of $Y_Q$. 
The freeze-out temperatures are nearly identical across $Y_Q$, as one would expect.
Generally, the values for $\left\{T^\text{FO},\mu_B^\text{FO}\right\}$ and $\left\{T^\text{FO},\mu_S^\text{FO}\right\}$ for different ion species overlap within the error bars for all $Y_Q$. 
The O+O predictions have significantly larger error bars due to  extrapolation issues. 
However, the results of $\left\{T^\text{FO},\mu_Q^\text{FO}\right\}$ do not all overlap at freeze-out. 
Because O+O collisions produce exactly symmetric nuclear matter, $\mu_Q=0$ by definition. 
We find that the ordering of $\mu_Q$ values––starting with the most negative: U+U, Au+Au, Zr+Zr, Ru+Ru, Cu+Cu, O+O––is consistent with their ordering in $Y_Q$, even though some error bars are still large. 

\begin{figure}[t!]
    \centering
    \includegraphics[width=0.85\columnwidth]{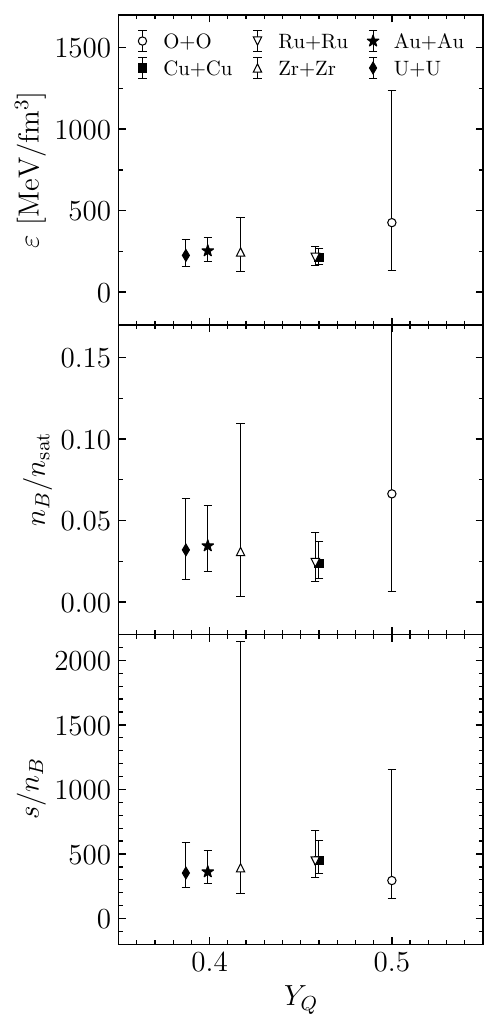}
     \caption{Charge fraction dependence of the energy density (top), baryon number density in terms of $n_\text{sat}$ (middle), and entropy per baryon number (bottom). Closed symbols are extracted from experimental data, open symbols are extracted from predictions for yield ratios.}
     \label{Fig:densities}
\end{figure}

Using the freeze-out parameters $\left\{T^\text{FO},\mu_B^\text{FO},\mu_S^\text{FO},\mu_Q^\text{FO}\right\}$, we can calculate the relevant thermodynamic quantities as functions of $Y_Q$. 
In Fig.~\ref{Fig:densities} we show the energy density versus $Y_Q$ (top), the baryon number density versus $Y_Q$ (middle), and the entropy per baryon number versus $Y_Q$ (bottom). 
Most relativistic hydrodynamic models choose a fixed energy density at freeze-out \cite{Plumberg:2024leb,Du:2023efk,Zhao:2020irc,Almaalol:2018gjh}, from our results here we find that $\varepsilon^\text{FO}(Y_Q) \approx \text{const.}$ such that a constant energy density approach is a good assumption at least across $Y_Q$.
Next, we compare the baryon density in terms of $n_\text{sat}$ where we assume $n_\text{sat}=0.16$ fm$^{-3}$. 
Given the current error bars in our results, $n_B(Y_Q)$ may either be constant or exhibit a slightly inverse scaling relation with $Y_Q$. 
However, systems with smaller $Y_Q$ also correspond to larger systems, increasing the probability of baryon stopping. 
With the given data we cannot delve deeper into this effect but leave it as a future topic when more experimental data becomes available. 
Finally, we observe the change in $s/n_B(Y_Q)$. 
In Ref.~\cite{Mroczek:2024sfp}, it was found using a relativistic mean-field model at a low constant temperature that $s/n_B$ decreases with decreasing $Y_Q$ (although $s/n_B(Y_Q)$ remained fairly flat for $Y_Q$ close to 0.5), which is consistent with our findings. 
What was most pertinent to constrain the development of 3-dimensional equations of state for neutron star mergers is the derivative shown in Eq.~(\ref{Eq:d2snB_ddelta2}) wherein we have rewritten $Y_Q$ in terms of $\delta$ defined in Eq.~(\ref{Eq:delta_definition}).
Using our results in Fig.~\ref{Fig:densities} (bottom), we fit polynomial functional forms to $s/n_B(Y_Q)$ within the error bars to compute the needed derivative.  
Because we require a second-order derivative, these functional forms must be at a least quadratic functions to obtain a non-zero result. Thus, we fit only quadratic and quartic forms to our results for $s/n_B(Y_Q)$.

\begin{figure}[htbp!]
    \centering
    \includegraphics[width=0.95\columnwidth]{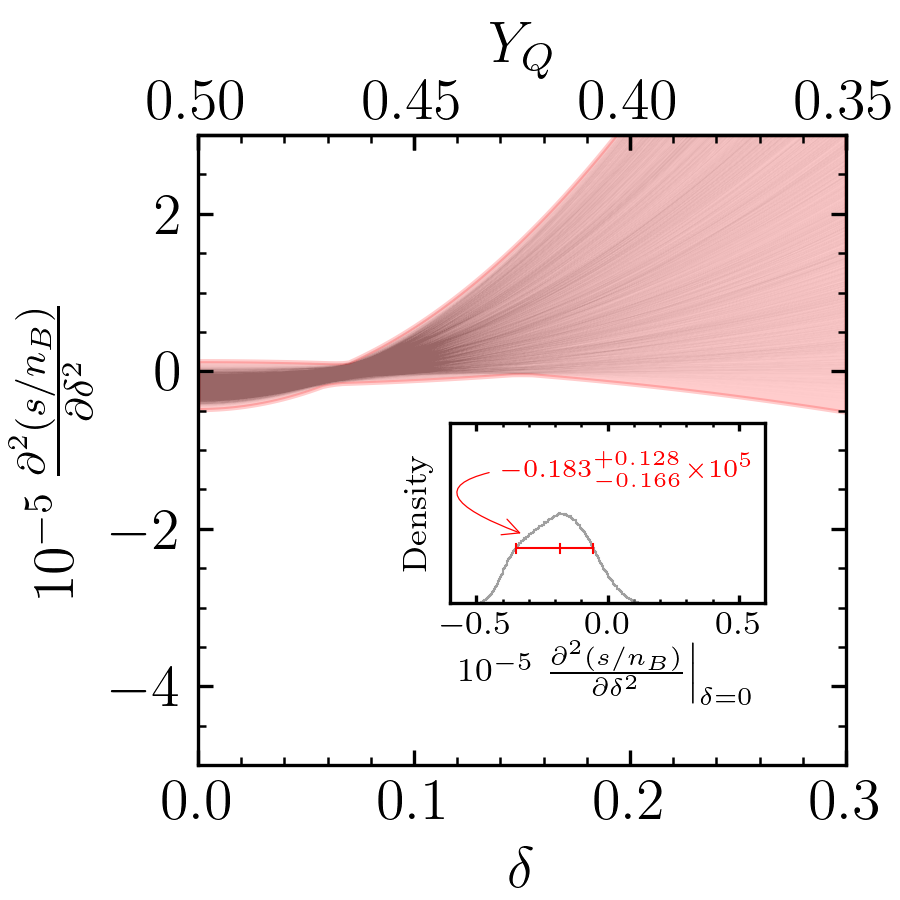}
    \caption{Second derivative of $s/n_B$ with respect to the isospin asymmetry parameter $\delta$ calculated from polynomial fits for the isentrope values extracted from thermal fits. The area bounding all possible quartic polynomials that fit data is shown in red, and a subset of the second derivative of these polynomials is depicted by gray lines. (Inset) Histogram of all resulting polynomials in the limit $\delta\rightarrow0$; result in red marks the value of the second derivative at the peak of the distribution and the uncertainty is fixed where the distribution falls $e^{-1/2}$ from the peak.}
    \label{Fig:d2snB_ddelta2}
\end{figure}

The results of $\partial^2 (s/n_B)/\partial \delta^2$ are shown in Fig.~\ref{Fig:d2snB_ddelta2} wherein we plot it versus $\delta$.  
This derivative should be taken at a $T=\text{const.}$, which is reasonable given Fig.~\ref{Fig:phase_diagrams} where $T^\text{FO}\approx 145$ MeV for all ion species considered.
For the neutron star merger equation of state, one would also want to take the $\delta\rightarrow 0$ limit, which is somewhat more challenging because our O+O results are predictions and not experimental data. 
Therefore, we argue that the O+O yield ratio results are crucial not only for constraining the hadron yield ratios but also for taking the $\delta\rightarrow 0$ limit.
Overall, the result of $\partial^2 (s/n_B)/\partial \delta^2$ is in qualitative agreement with what was found in Fig. 6 of Ref.~\cite{Mroczek:2024sfp}.
However, ideally, we would have results across multiple different $T$, $n_B$, and $\delta$ to reconstruct this derivative across a wider region of the phase diagram. 

{\it Conclusions and outlook.}––In this paper we use existing hadron yield ratios at top RHIC energies to make predictions for the yield ratios of O+O, Ru+Ru, and Zr+Zr.  Then we extract thermodynamic quantities at freeze-out across a range in $Y_Q$. 
From our analysis, we are able to provide new insights into the effect of $Y_Q$ on the equation of state. 
We find that the assumption of $\varepsilon=\text{const.}$ at freeze-out appears to be a good assumption as one changes $Y_Q=Z/A$ of the colliding ions. 
We find that the ratio of $s/n_B$ increases with $Y_Q$ such that if one calculates the second derivative of $s/n_B$ at fixed $T$ with respect to $\delta$ we obtain a negative derivative, which is consistent with theoretical calculations \cite{Alford:2022bpp} in the regime relevant to neutron star mergers \cite{Mroczek:2024sfp}.
Finally, we find that experimental hadron yield results for O+O would be the most informative because we require experimental data for symmetric nuclear matter.

Our results demonstrate the importance of having hadron yield ratio data across a wide range of $Y_Q=Z/A$ (different choice of colliding species) and different $T$ and $\mu_B$ (different choice of $\sqrt{s_{NN}}$). 
Potential ions of interest that are both stable and provide useful $Y_Q$ values are: $^3$He with $Y_Q\approx0.66$, $^{7}$Be with $Y_Q\approx0.57$, $^{40}$Ca or $^{44}$Ti with $Y_Q=0.50$, $^{58}$Ni with $Y_Q\approx0.483$, and $^{96}$Mo of $Y_Q\approx0.434$.
Another interesting possibility to reach even lower values in $Y_Q$ is the use unstable beams.
Further runs using distinct colliding ion species across different $\sqrt{s_{NN}}$ would allow us to calculate the second derivative of $s/n_B$ with respect to $\delta$, essential to constraining the equation of state of neutron star mergers.
Low collision energies would be ideal to access information at large $\mu_B$ and lower $T$. 
While the STAR experiment has already finished its Beam Energy Scan runs, there is still an opportunity to run scans in $Y_Q$ and $\sqrt{s_{NN}}$ in the upcoming 
Compressed Baryonic Matter (CBM) at FAIR \cite{Almaalol:2022xwv}. 
Additionally, it would also be interesting to conduct a similar $Y_Q$ scan at the LHC, which will eventually be possible using $^{208}$Pb, $^{129}$Xe, $^{16}$O that translate to $Y_Q=0.39,0.42,0.5$, respectively. 
Combining our results here with an LHC scan in $Y_Q$ would then allow us to construct the $\partial^2 (s/n_B)/\partial \delta^2$ derivatives at two different points in $\mu_B$ for a nearly constant $T$.  
Future $Y_Q$ scans could also provide insight to the kaon isospin anomaly \cite{ALICE:2021fpb,NA61SHINE:2023azp,Brylinski:2023nrb} and the disorientated isospin condensate \cite{Gavin:2001uk,Kapusta:2023xrw}.

{\it Acknowledgements.}––The authors would like to thank D. Mroczek for useful discussions during the preparation of this manuscript. The authors also thank P. Tribedy, R. Ma, and C.Y. Tsang for their comments on recent experimental data. The authors acknowledge support from the US-DOE Nuclear Science Grant No. DE-SC0023861, and within the framework of the Saturated Glue 34 (SURGE) Topical Theory Collaboration. 
This work was also supported in part by the National Science Foundation (NSF) within the framework of the MUSES collaboration, under grant number OAC-2103680 and from the Illinois Campus Cluster, a computing resource that is operated by the Illinois Campus Cluster Program (ICCP) in conjunction with the National Center for Supercomputing Applications (NCSA), and which is supported by funds from the University of Illinois at Urbana-Champaign.

\bibliography{inspire,not-inspire}

\appendix

\section{SUPPLEMENTAL MATERIAL}

{\it Thermal Fitting procedure.}––In describing the procedure used in our fit, we obtain the best model fit by minimizing the distribution of $\chi^2$,
\begin{equation}
\chi^2\equiv\sum_i \frac{\left(N_i^{\text {Exp }}-N_i^{\text {HRG}}\right)^2}{\sigma_i^2},
\end{equation}
and obtaining an estimate of the systematic error by considering the quadratic deviation $\Delta^2$ defined as:
\begin{equation}
\Delta^2\equiv\sum_i \frac{\left(N_i^{\text {Exp }}-N_i^{\text {HRG }}\right)^2}{\left(N_i^{\text {HRG }}\right)^2}.
\end{equation}
where $N_i^\text{Exp}$ and $N_i^\text{HRG}$ are the experimentally measured value and HRG model calculation of the $i$-th ratio yield, respectively, with its uncertainty denoted by $\sigma_i$. For the experimental errors, we added the systematic errors of the measured ratios; for the predicted error, we added the upper bound error of the model calculation. The sum runs over all ratios. The best-fitting chemical freeze-out parameters $\left\{T^\text{FO},\mu_B^\text{FO},\mu_S^\text{FO},\mu_Q^\text{FO}\right\}$ are then extracted from this $\chi^2$-minimization procedure. 

\begin{figure}[htpb!]
    \centering
    \includegraphics[width=0.95\columnwidth]{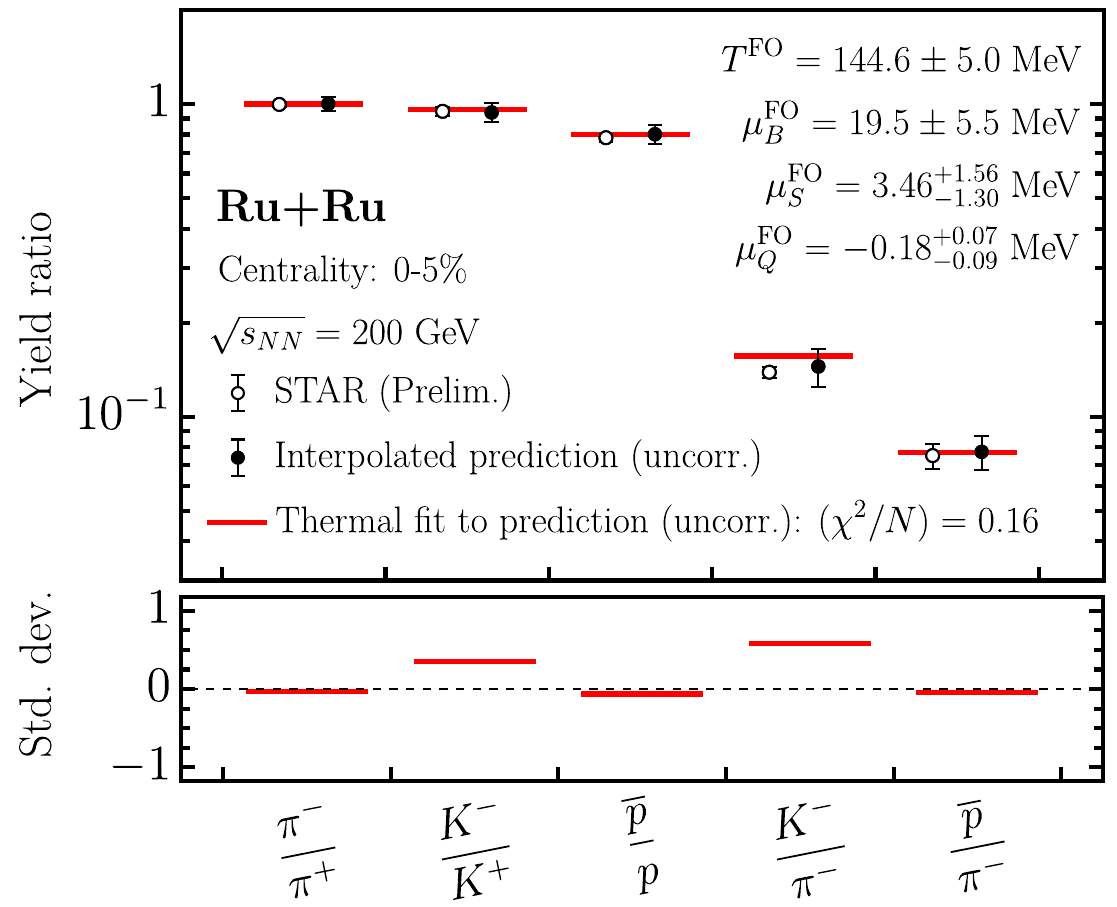}
    \caption{Thermal fit of hadron yield ratios of identified charged particles for Ru+Ru at $\sqrt{s_{NN}}=200$ GeV collisions obtained from interpolation of experimental values of other collision systems (see main text). The extracted yield ratios presented in Table~\ref{Table:predicted_yield_ratios} are represented by closed circles. Preliminary data from STAR measurements taken from Ref.~\cite{Ma:2024wwnd} is represented by open circles and displayed only as reference.
    }
    \label{Fig:thermal-fit_RuRu}
\end{figure}

\begin{figure}[htpb!]
    \centering
    \includegraphics[width=0.95\columnwidth]{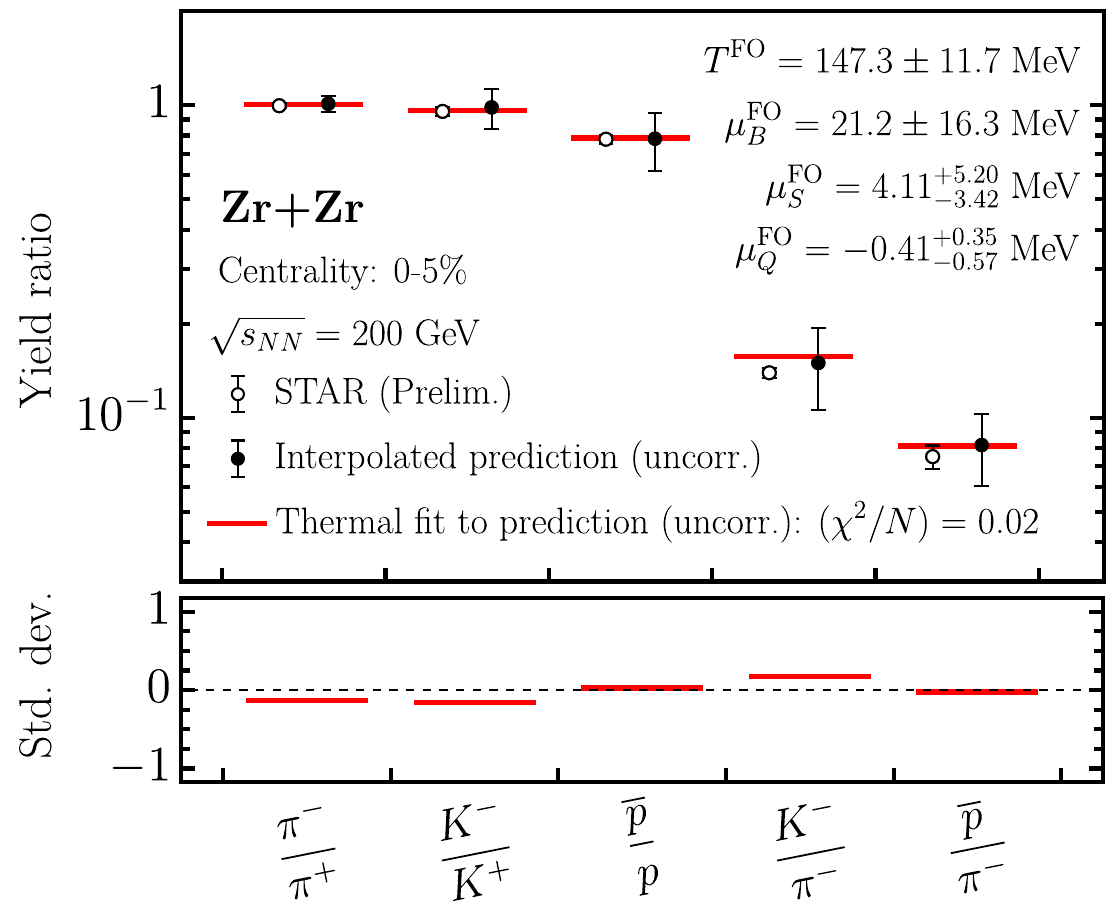}
    \caption{Thermal fit of hadron yield ratios of identified charged particles for Zr+Zr at $\sqrt{s_{NN}}=200$ GeV collisions obtained from interpolation of experimental values of other collision systems (see main text). The extracted yield ratios presented in Table~\ref{Table:predicted_yield_ratios} are represented by closed circles. Preliminary data from STAR measurements taken from Ref.~\cite{Ma:2024wwnd} is represented by open circles and displayed only as reference.
    }
    \label{Fig:thermal-fit_ZrZr}
\end{figure}

Figures \ref{Fig:thermal-fit_RuRu} and \ref{Fig:thermal-fit_ZrZr} show the results of the thermal-fitting procedure applied to the $\pi^-/\pi^+$, $K^-/K^+$, $\bar{p}/p$, $K^-/\pi^-$, and $\bar{p}/\pi^-$ hadron yield ratios for Ru+Ru and Zr+Zr, respectively. In both cases, red lines indicate the yield ratios produced by the hadron resonance gas with chemical freeze-out parameters $\{T^\text{FO},\mu_B^\text{FO},\mu_S^\text{FO},\mu_Q^\text{FO}\}$ resulting from the fit to the yield ratios represented by closed circles. Preliminary STAR measurements \cite{Ma:2024wwnd} are shown as reference.

{\it Parameter table.}––The coefficients of the first-, second-, and third-order polynomial functions $p_m(Y_Q)$, used to interpolate and extrapolate the hadron yield ratios as functions of $Y_Q$ in Fig.~\ref{Fig:predicted_yield_ratios}, can, in principle, vary widely. In our analysis, different hadron yield ratios required different sets of parameters. We selected these parameters to cover the largest possible parameter space while keeping the computational effort reasonable. A summary of the chosen parameters is presented in Table~\ref{Table:YQ_polynomial_coefficients}.

\begin{widetext}

\begin{table*}[htbp!] 
\centering
\begin{tabular}{l@{\hspace{30pt}}c@{\hspace{10pt}}c@{\hspace{30pt}}c@{\hspace{10pt}}c@{\hspace{30pt}}c@{\hspace{10pt}}c}
    \toprule
    Ratio/parameter & $a_{1,i}$         & $\Delta_{1,i}$    & $a_{2,i}$             & $\Delta_{2,i}$    & $a_{3,i}$         & $\Delta_{3,i}$    \\
    \colrule
    $\pi^-/\pi^+$   & $[-4.0, 4.0]$     & 0.001             & $[-50.0, 50.0]$       & 0.50              & $[-50.0, 50.0]$   & 1.0               \\
    $K^-/K^+$       & $[-10.0, 10.0]$   & 0.010             & $[-150.0, 150.0]$     & 0.10              & $[-100.0, 100.0]$ & 0.5               \\
    $\bar{p}/p$     & $[-10.0, 10.0]$   & 0.010             & $[-150.0, 150.0]$     & 0.10              & $[-100.0, 100.0]$ & 0.5               \\
    $K^-/\pi^-$     & $[-10.0, 10.0]$   & 0.010             & $[-150.0, 150.0]$     & 0.10              & $[-100.0, 100.0]$ & 0.5               \\
    $\bar{p}/\pi^-$ & $[-10.0, 10.0]$   & 0.010             & $[-150.0, 150.0]$     & 0.05              & $[-100.0, 100.0]$ & 0.5               \\
    \botrule
\end{tabular}
\caption{Ranges for coefficients $a_{m,i}$ and spacings $\Delta_{m,i}$ used for the estimation of all possible polynomials $p_m(Y_Q)$ (where $m=\{1,2,3\}$) that satisfy experimental data points and uncertainties for all hadron yield ratios considered. All coefficients of $p_m(Y_Q)$ have the same parameter range for fixed order, i.e., independently of $i=\{0,\ldots,m\}$.
}
\label{Table:YQ_polynomial_coefficients}
\end{table*}
\end{widetext}

\end{document}